\documentclass[11pt]{article}
\usepackage{hyperref}
\pdfoutput=1
\begin{document}
\title{The Lagrangian picture of heat transfer in convective turbulence}
\author{Maik Boltes$^1$, Herwig Zilken$^1$ and J\"org Schumacher$^2$ \\ \\
$^1$ J\"ulich Supercomputing Centre, Research Centre J\"ulich,\\D-55425 J\"ulich, Germany,\\
$^2$ Institute of Thermodynamics and Fluid Mechanics, \\Ilmenau University of Technology,
D-98684 Ilmenau, Germany}
\maketitle

\begin{abstract}
We present a fluid dynamics video which illustrates the Lagrangian aspects of local heat transfer 
in turbulent Rayleigh-B\'{e}nard convection. The data are obtained from a direct numerical simulation.
\end{abstract}

\section{Video Description}
If a fluid in an extended layer or a closed cell of height $H$ is cooled from above and heated from below turbulent 
convection is initiated when the temperature drop between top and bottom plates is sufficiently
large. The two videos,  \href{http://ecommons.library.cornell.edu/handle/1813/13639}{Video1} and 
\href{http://ecommons.library.cornell.edu/handle/1813/13639}{Video2} (in higher resolution), show the motion of Lagrangian tracer particles in 
a Cartesian turbulent  convection cell with periodic side walls and free-slip boundary 
conditions at the bottom and top plates ($z=0$ and $H$). The direct numerical simulations solve the 
Boussinesq equations of Rayleigh-B\'{e}nard convection and are based on a
pseudospectral method. The computational grid contains $N_x\times  N_y\times N_z=2048\times 2048\times 513$ 
points. A set of 1.2 million tracers is advected with the flow. The simulation parameters
are Prandtl number $Pr=0.7$, Rayleigh number $Ra=1.2\times 10^8$ and aspect ratio 
$\Gamma=L/H=4$ with $L$ being the sidelength in $x$ and $y$ directions \cite{Schumacher2008,Schumacher2009}. 

The subset of Lagrangian tracers, which is shown in the first part of the video, starts in a
plane right above the thermal boundary layer. The tracer particles are colored with respect 
to the total temperature $T$ (red=hot and blue=cold) at their position. The diagonal cut through the 
evolving particle 
cloud illustrates the thermal plumes, which carry locally heat through the layer.  A number 
of selected tracers are shown together with their particle track.

Heat is carried through the layer convectively when the product of vertical velocity fluctuation
and temperature fluctuation is positive, $u_z\theta>0$. This can be both, a hot rising plume and a cold downwelling
plume.  The final part of the animation colors the tracers with respect to their amplitude of 
$u_z\theta$. Levels of red stand for $u_z\theta>0$ and white for $u_z\theta\le 0$.

The computations have been conducted on 4096 cores of the Blue Gene/P JUGENE at the 
J\"ulich Supercomputing Centre under grant HIL02. JS acknowledges support by the Deutsche 
Forschungsgemeinschaft within the Heisenberg Program.

\end{document}